# Apparatus Design for Measuring of the Strain Dependence of the Seebeck Coefficient of Single Crystals

Tiema Qian,[1*] Joshua Mutch,[1*] Lihua Wu,[2] Preston Went,[1] Jihui Yang,[2] Jiun-Haw Chu[1]

[1]*Department of Physics, University of Washington, Seattle, Washington, 98105, USA*

[2]*Materials Science and Engineering, University of Washington, Seattle, Washington, 98105, USA*

*The above authors contributed equally to this work

We present the design and construction of an apparatus that measures the Seebeck coefficient of single crystals under in-situ tunable strain at cryogenic temperatures. A home-built three piezostack apparatus applies uni-axial stress to a single crystalline sample and modulates anisotropic strain up to 0.7%. An alternating heater system and cernox sensor thermometry measures the Seebeck coefficient along the uniaxial stress direction. To demonstrate the efficacy of this apparatus, we applied uniaxial stress to detwin single crystals of $BaFe_2As_2$ in the orthorhombic phase. The obtained Seebeck coefficient anisotropy is in good agreement with previous measurements using a mechanical clamp.

**I. INTRODUCTION**

The Seebeck coefficient is the ratio of the voltage difference to the temperature gradient in a material:

$$S = -\frac{\Delta V}{\Delta T}$$

Measurements of the Seebeck coefficient provide a uniquely different and complementary probe compared to resistivity measurements, in part because of its sensitivity to particle-hole asymmetry[1]. For example, it is often used to determine if a semiconductor is *p*-type or *n*-type simply by the sign of the coefficient. The sign, magnitude and anisotropy of Seebeck coefficient provide strong constraint to the theory of strongly correlated materials. Additionally, searching for materials with a large Seebeck coefficient is an active area of research due to applications in power generation, thermometry, and electronic refrigeration [2-4].

In the past few years, in-situ tunable strain has proven to be a powerful tool to probe and control exotic phases in both topological [5, 6] and strongly correlated materials[7-10]. Most of the in-situ strain work to date relies on the measurement of electrical resistivity as a probe of the electronic structures. In this paper, we introduce an apparatus design that measures the Seebeck coefficient as a function of in-situ strain. The apparatus described here offers in-situ control of strain up to +/- 0.7% strain at 100K, while simultaneously measuring the Seebeck coefficient with fine resolution. As a demonstration, we performed measurements on single crystal of $BaFe_2As_2$. $BeFe_2As_2$ goes through a tetragonal to orthorhombic structural transition near 135K and forms dense twin domains. We applied uniaxial stress to detwin the crystal and determined the Seebeck coefficient anisotropy, which is in agreement with previous measurements using a mechanical clamp[11, 12].

## II. Experimental Setup

We used a home-built three-piezo strain apparatus to control strain, as shown in Fig. 1a. This three-piezo technique was first introduced by Hicks et al[13]. It has the advantage of minimizing uncontrolled strain bias due to the mismatch in thermal contraction. This apparatus also provides a large in-situ uni-axial strain tunability. As shown in Fig. 1, the apparatus induces tensile and compressive stress by changing the width of a gap between two pieces of titanium blocks, which are glued to the outer and inner piezostacks, respectively. Applying voltages of opposite polarity to the outer and the inner piezostacks results in a displacement of titanium blocks, which applies a uni-axial stress to a crystal glued across the gap. For the thermopower measurement, a 0.031-inch thick piece of G-10 was glued on either side of this gap to provide thermal insulation. Next, cernox sensors (CX-1050-SD, Lake Shore Cryotronics) were glued on top of this G-10 layer. Finally, strain-gauge heaters (KFH-1.5-120-C1-11L1M2R, OMEGA) were glued on each cernox sensor, and a crystal was glued across the gap between the cernox sensors, as shown in Fig. 1b and 1c. We used STYCAST 2850FT, Loctite epoxy for each layer of glue. A strain gauge (MMF003096, Micro-measurements) was glued on the side surface of one of the outer piezostacks to measure its strain, $\epsilon_{piezo}$. The total strain induced by the gap displacement is estimated as $\epsilon_{xx}^{disp} = 2 \times \frac{L}{l} \times \epsilon_{piezo}$, where $L$ is the length of the piezostack (9mm) and $l$ is the adjustable width of the apparatus gap. For the apparatus presented in this paper $l\sim$0.75mm, which translates to $\epsilon_{xx}^{disp} \sim 0.7\%$ at T = 100K for typical voltages applied to the piezostacks. The actual strain experienced by the crystal is typically 70% of $\epsilon_{xx}^{disp}$ due to the strain relaxation in the G10, epoxy and cernox. A detailed finite element simulation of strain relaxation is presented in the appendix.

We used a Janis flow cryostat (ST-100, Janis Research) to provide the cryogenic environment. Since the Janis cryostat operates under vacuum pressures of order $\sim$10$^{-6}$ Torr, heat conduction from the cold head to the sample must be bridged by the piezostacks if no other conduction path is provided. The piezostacks are extremely thermally insulating. In order to more efficiently cool the sample, we used twisted bare copper wire pairs to provide addition heat conduction path. As shown in Fig. 1a, the twisted wires were adhered to the titanium blocks by STYCAST 2850FT epoxy and the wires are flexible enough to allow the displacement of titanium blocks.

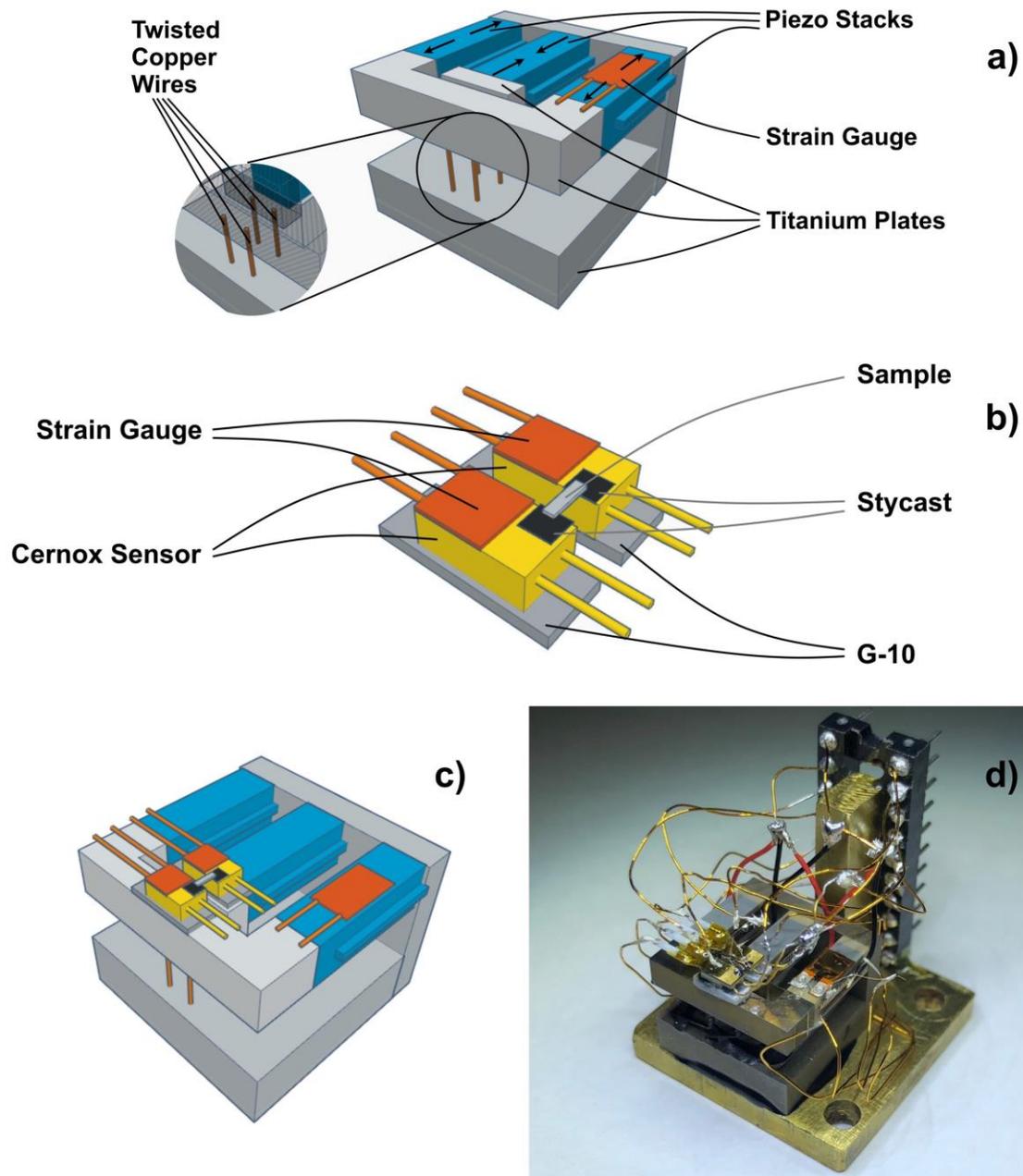

FIG.1. Schematic diagrams and photograph of the three piezostack strain apparatus for the Seebeck measurement. (a) Three piezostack strain apparatus: Piezostacks are shown blue, strain gauges in orange, titanium plates in gray, and the twisted copper thermal bridges are shown in the inset. Arrows show the displacement of the piezostacks when providing tensile strain. (b) Seebeck measurement components. The strain gauges are shown orange. The two smaller strain gauges act as heaters, while the larger strain gauge measures $\epsilon_{piezo}$. The cernox sensors are shown in yellow, the G-10 plates are shown in dark gray, and the STYCAST used to adhere the crystal is shown in black. (c) The assembled entire apparatus, combining both the strain mechanics shown in (a) and the Seebeck components shown in (b). The Seebeck measurement components are mounted across the gap that is displaced by the strain apparatus. (d) Photograph of the apparatus.

The electrical schematic of the apparatus is shown in Fig. 2, following the design introduced by Eundeok et al[14]. To minimize unwanted emfs arising from thermal gradients in the cryostat, we used phosphor bronze wiring for the sample voltage measurement. A 32-gauge phosphor bronze wire (Lake Shore Cryotronics) was used to connect the electrical feedthrough of the Janis to the top of the cold head. This wire was briefly interrupted by a stainless-steel dip socket connection for ease of assembly, and eventually connected to 25-micron diameter phosphor bronze wire (Goodfellow, 346-032-19). This 25-micron diameter phosphor bronze wire made contact to the crystal using DuPont 4929 silver paint. To further eliminate an offset error resulting from thermal emfs, we employed a "heater switching" method, similar to a method reported by Eundeok et al [14]. By alternating heating each end of the sample, erroneous thermal emfs in the circuitry can be eliminated by examining the difference of the sample voltage response to each heating configuration.

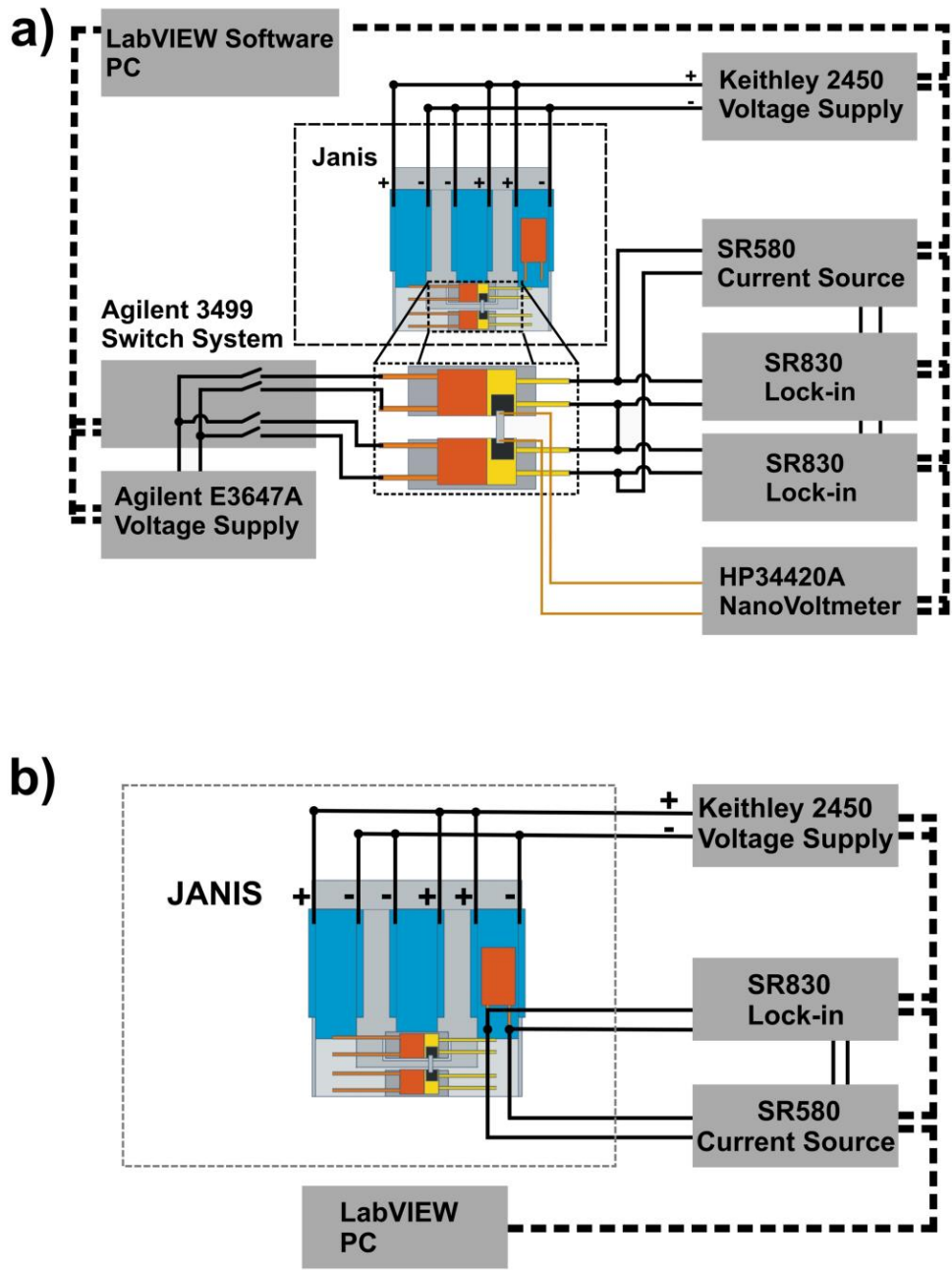

FIG. 2: Block diagram of the experimental circuitry for (a) Seebeck measurement and (b) strain measurement. (a) The system temperature is controlled by a Lakeshore 335 temperature controller. All instruments shown are controlled by LabVIEW software via a GPIB interface. The strain gauges acting as heaters are powered by a DC voltage supply (E3647A, Agilent). The resistances of the cernox sensors are measured by two Lock-in Amplifiers (SR830, Stanford Research System) and a voltage controlled current source (CS580, Stanford Research System). The voltage applied to the parallel piezo stacks is controlled by an interactive source meter (Model 2450, Keithley). The sample voltage is measured by a nanovoltmeter (HP34420A). (b) Circuit diagram for the strain measurement of $\epsilon_{piezo}$. The resistance of the strain gauge is measured using a 4-point measurement with one of the lock-in amplifiers and the voltage controlled current source. Altogether, the system uses 18 electrical pins, making it possible to do with a standard 19-pin feedthrough.

A single crystalline sample of BaFe$_2$As$_2$ was prepared by cleaving and cutting into a rectangular shape. The dimensions of the sample are approximately $1mm \times 0.1mm \times 0.01mm$. The sample was sputtered with gold on two ends and a two points contact was made using phosphor bronze wire and silver paste.

### III. Measurement Procedure

Prior to the measurement, the cernox censors were calibrated in a in order to obtain an accurate temperature profile on both sides of the sample. The setup is the same as shown in Fig. 2a, except the sample voltage wasn't recorded and the heater voltage was set to 0V. The resistance of both cernox sensors was measured at temperature set points in 1.5K intervals. To overcome any thermal lag, we waited for 10 minutes for the apparatus to reach temperature stability at each temperature setpoint. This data was linearly interpolated to provide a temperature calibration over the relevant temperature range.

The Seebeck coefficient was measured at fixed strain setpoints. First, the cryostat was ramped to a temperature setpoint, then given 30 minutes to establish temperature stability. Prior to performing the Seebeck measurement, a strain measurement of $\epsilon_{piezo}$ was performed to estimate the strain delivered to the crystal, using the circuitry shown in Fig. 2b. A voltage triangle wave was applied to the piezostacks, as shown in Fig. 3a. This triangle wave looped three times to inspect the repeatability of the piezostack hysteresis curve. At the same time, the resistance of the foil strain gauge glued to one of the piezostacks was measured to determine $\epsilon_{piezo}$, shown in Fig. 3b. The offset expansion feature of the SRS 830 Lock-in was used to enhance the sensitivity of the resistance measurement. By applying this voltage waveform to the piezostacks prior to measuring the Seebeck coefficient, the piezostacks are trained to operate on a consistent hysteresis curve, and the measured $\epsilon_{piezo}$ can be used to calibrate the Seebeck measurement later.

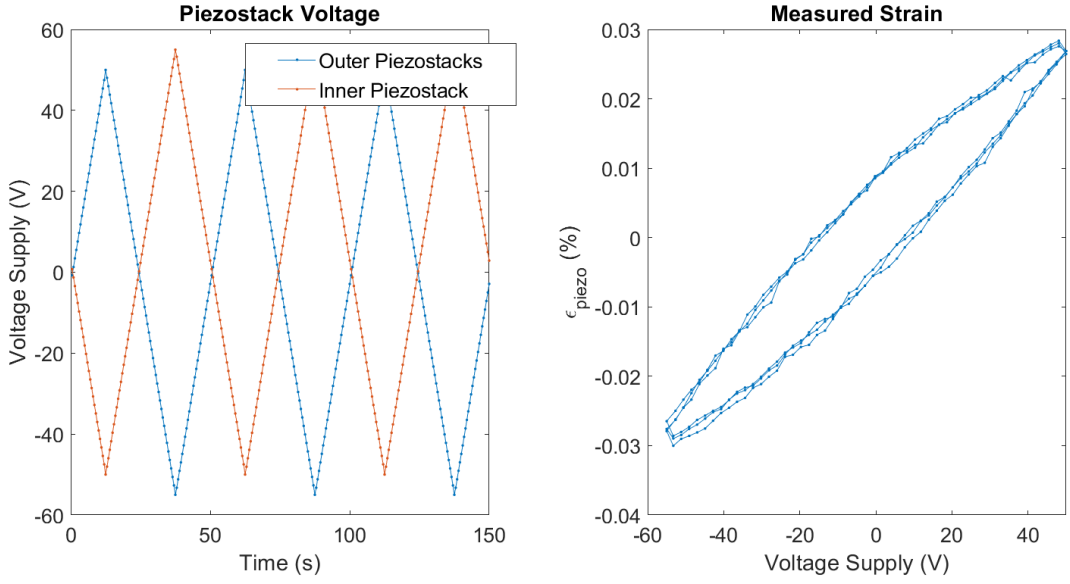

FIG. 3. Strain provided by piezostacks versus voltage supply to piezostacks at 120K (a) Voltage across the inner and outer piezostacks with respect to time. (b) $\epsilon_{piezo}$ measured by the foil strain gauge versus voltage supply across one of the outer piezostacks. Three voltage loops across the piezostacks were performed in order to inspect the repeatability of the piezostack hysteresis.

After training the piezostacks and measuring $\epsilon_{piezo}$, the voltage of the piezostacks was ramped to a fixed setpoint, and given one minute to stabilize. Next, one heater (Heater #1) attached to one of the cernox sensors was turned on by applying a DC voltage across the heater. After two minutes, Heater #1 was turned off, and Heater #2 was turned on. Heater #2 stayed on for two minutes, then was turned off as well. Both heaters were kept off for 1 minute to complete one heating cycle. During this heating cycle, the temperature of each cernox sensor and the voltage output from the sample was recorded and displayed live time by a LabVIEW program, similar to what is shown in Fig. 4A and 4B. Monitoring this output in real time allowed the parameters (heating time, heating power) to be adjusted as necessary to ensure that a stable temperature gradient is reached within the heating cycle. We targeted $\Delta T = \sim 1K - 2K$, and found heating powers of $3mW - 15mW$ generally accomplished this. After completing one heating cycle, the voltage across the piezostacks was then ramped to the next set point, and another heating cycle began. The strain loop was repeated at least three times to inspect the repeatability of the result.

The Seebeck coefficient was determined by using the difference in temperature and difference in sample voltage extracted from the equilibrium portions of the dataset, after each heater had been on for two minutes. We follow the explanation given in Ref. 14 to extract the Seebeck coefficient. Letting the subscript $i$ indicate the time just before alternating power to the heaters, and the subscript $f$ indicate the time just before turning

off both the heaters, then

$$2\Delta T = T_{2f} - T_{1f} + (T_{1i} - T_{2i})$$

$$2\Delta V = V_f - V_i$$

$$T_{avg} = \frac{T_{1f} + T_{1i} + T_{2f} + T_{2i}}{4}$$

$$S = -\frac{2\Delta T}{2\Delta V}$$

It is recognized that both the factor of 2 in $2\Delta T$ and $2\Delta V$ results from the alternative heating scheme. The temperature corresponding to the measured Seebeck coefficient corresponds most closely to $T_{avg}$.

From the data recorded during each heating cycle, the Seebeck coefficient $S$ was plotted against the piezostack voltage, as shown in Fig. 4d. The pronounced hysteresis is associated with the hysteretic strain-voltage response of piezostacks. To remove this hysteresis effect, the measured strain $\epsilon_{piezo}$ from the foil strain gauge was used to calibrate strain per voltage for increasing and decreasing voltage. By estimating $\epsilon_{xx}^{disp} = 2 \times \frac{L}{l} \times \epsilon_{piezo}$, as discussed in the Experimental Setup section, the Seebeck coefficient can be plotted against the apparatus strain, and the hysteresis effect is removed. Our finite element analysis (see appendix) estimates a relaxation factor $\alpha \sim 0.67$, which indicates that the strain delivered to the sample is $\epsilon_{aa} = \alpha \epsilon_{xx}^{disp}$. In Fig.5, we plot the Seebeck coefficient versus $\epsilon_{aa}$ for the same dataset as shown in Fig. 4d, as well as an additional dataset that includes higher magnitude strains. A large Seebeck coefficient response to strain is seen for low strains, with a saturating behavior at large strains. The saturated values are consistent with previously reported values [11, 12] measured by a mechanical clamping method. We have also measured the Seebeck coefficient of an unstrained crystal using a Quantum Design thermal transport option (TTO). The temperature dependence of Seebeck coefficient of the stress-free crystal is shown in the inset of Fig. 5, which is in good agreement with previous reports [15-19].

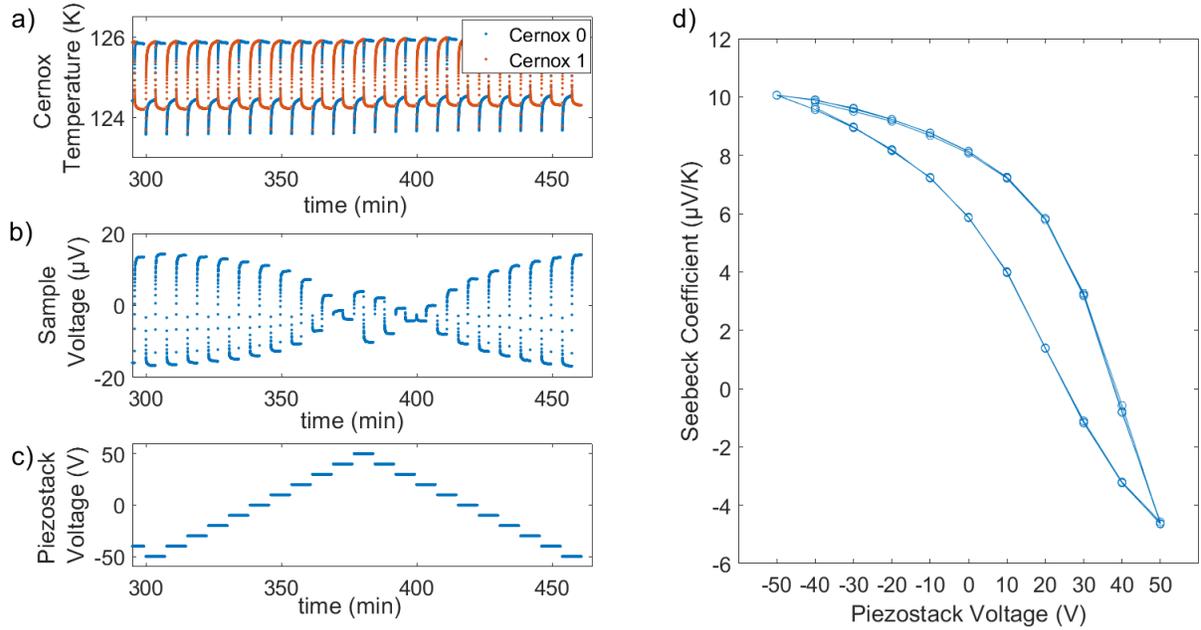

FIG. 4 Data taken on our home-built apparatus measuring the Seebeck coefficient response to strain for the test material BaFe$_2$As$_2$ at 125K. (a) Temperature on each side of the sample versus time. Each side of the sample was heated in an alternating fashion. (b) Voltage measurement across the sample for approximately 20 heat cycles in one strain loop (c) Voltage supplied to the piezostacks. Positive (negative) voltage across the piezostacks corresponds to tensile (compressive) stress applied to the sample. The time scale of the x axis in (a-c) corresponds to one strain loop (d) Calculated Seebeck coefficient versus piezostack voltage. Several strain loops of data are shown to inspect repeatability of the measurement. A hysteresis is seen, which is associated with the hysteretic behavior of the piezostacks.

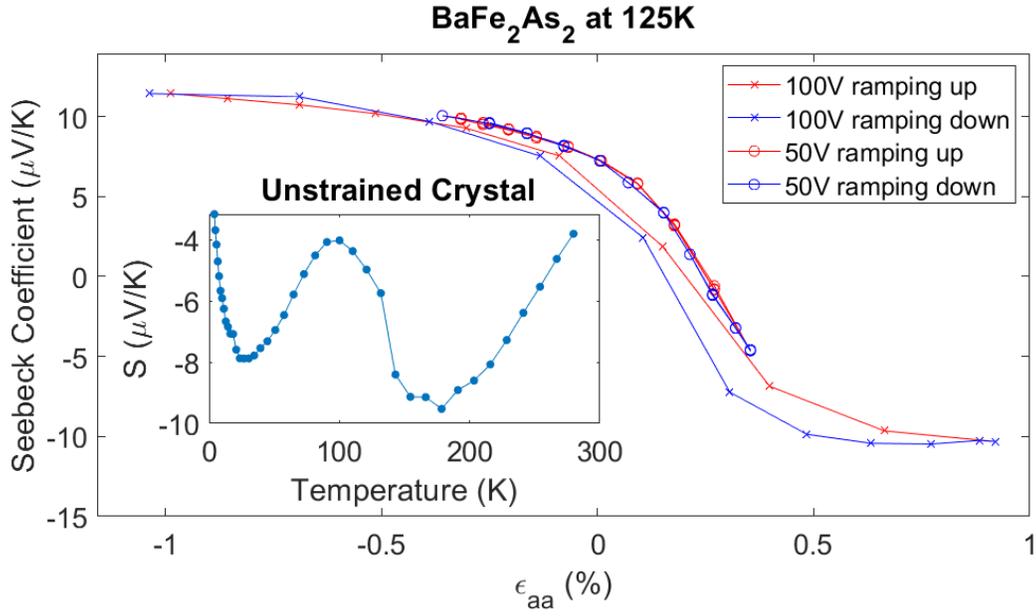

FIG. 5 The Seebeck Coefficient verses strain of BaFe2As2 at 125K. Increasing strain and decreasing strain are shown in two different colors. The hysteresis seen in Fig. 4d was removed by using $\epsilon_{piezo}$ as measured by the strain gauge for increasing and decreasing strain. The strain reported in the x axis, $\epsilon_{aa}$, is given by $\epsilon_{aa} = 0.67 \times 2 \times \frac{L}{l} \times \epsilon_{piezo}$, as described in the Experimental Setup section. where 0.67 is a relaxation factor calculated by finite element analysis (see appendix). INSET: The Seebeck coefficient of unstrained BaFe2As2, as measured by a Quantum Design thermal transport option (TTO) along the Fe-Fe bonding direction of the crystal. The temperature dependence agrees qualitatively with previously reported measurements

## IV. Discussion and Conclusion

Most of the strain dependent Seebeck coefficient measurements reported to date have been done by mechanical clamping [11, 12]. While this technique has the benefit of delivering large amounts of strain, controllable in-situ strain measurements are desirable to monitor the exact Seebeck-strain dependence. A very recent report demonstrated the measurements of elasto-Seebeck and elasto-Nernst measurements of the 1111 iron-based superconductors by gluing the crystals on the sidewall of piezo-stacks, yet the amount of strain that can be delivered by this method is an order of magnitude smaller than the our setup [20]. Although we only present measurements in zero field, the set up could be easily extended to an environment with magnetic fields because of the low field dependence of the cernox thermometry used. Field-dependent thermoelectric measurements have already provided useful information in strongly correlated electron systems and high-Tc superconductors. Adding another control knob of strain can be a useful and exciting tool.

**Appendix. Finite Element Analysis**

In order to estimate the strain transmission, defined as the ratio of strain delivered to the crystal to the apparatus strain, we modelled our setup with the *ANSYS Academic Research Mechanical 19.1* finite element analysis package. 3-piezo strain apparatuses such as the one we use have been used before in literature[5, 6, 13]. In these simpler systems, the strain transmission is generally above 70%, depending on the crystal composition and dimensions and mounting technique. However, in our more complicated system, G-10, Cernox sensors, and finally the crystal are stacked together and adhered with stycast epoxy. This allows for a stronger relaxation effect associated with the glue. Our finite element model investigates if these effects are significant.

We modelled a 2mm x 0.2mm x 0.02mm $BaFe_2As_2$ crystal glued across our apparatus. The elastic coefficients of this crystal were sourced from Fujii et al[21]. at 250K, with the missing $C_{13}$ coefficient estimated as 34 GPa by comparison between Fujii et al. and the stiffness tensor calculated by the Materials Project [22]. The crystal was mounted such that strain was applied along the Fe-Fe direction. The cernox sensors, stycast glue, and G-10 plastic were all modelled as isotropic, and the values for the Young's modulus and Poisson's ratio used in our model are given in Table 1. The stycast glue is modelled as 0.02mm thick for most connections, and 0.01mm thick between the cernox sensor and the crystal.

| Component | Young's Modulus (GPa) | Poisson's ratio |
|---|---|---|
| Cernox sensor | 400 | 0.245 |
| G-10 | 18 | 0.13 |
| Stycast epoxy[13] | 15 | 0.3 |

Table 1: Physical properties used in the finite element analysis model. The cernox sensors were modelled as 99.5% alumina. While the physical properties of alumina depend on purity, the Young's modulus is always at least one order of magnitude higher than any other material in our apparatus. The G-10 properties are were modelled according to the manufacturer datasheet.

In our model, we analyzed a 0.1% tensile strain applied to the apparatus and calculated the average strain transmission in the crystal in the strained region. This yielded a strain transmission of 67%, as opposed to 88% with the same crystal similarly epoxied to a bare 3-piezo device. Fig. 6 shows a schematic of the simulation results.

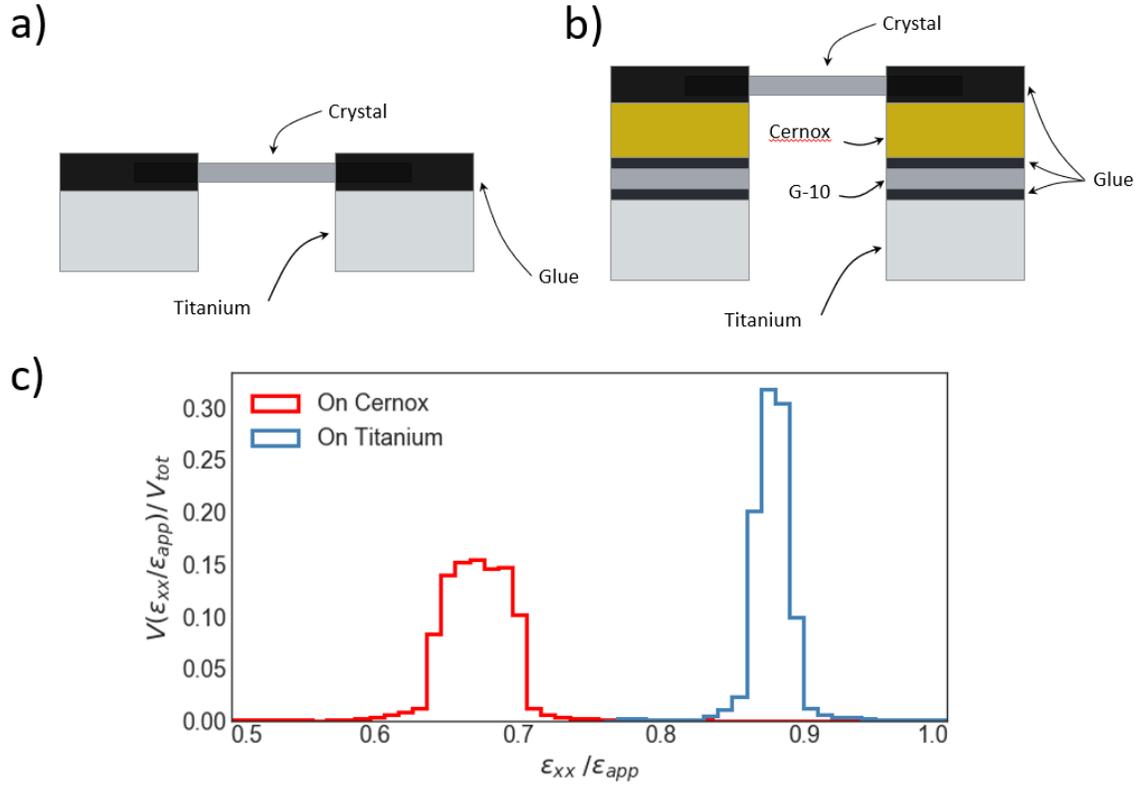

FIG. 7 a) Schematic (not to scale) of the Finite Element Analysis (FEA) model for a crystal glued directly across two titanium block in a simple strain apparatus. The gap between the titanium blocks was displaced to induce a strain of $\epsilon_{xx}^{disp} = 0.1\%$. b) Schematic of the FEA model for a crystal glued to our strain and Seebeck apparatus. Several elements are stacked together and adhered with glue, allowing for a stronger relaxation effect associated with the glue. The glue thickness between the apparatus and G-10, and between the G-10 and Cernox sensor was modelled as 0.02mm thick. The glue thickness between the Cernox sensor and crystal was modelled as 0.01mm thick. c) Average strain transmission for models shown in a) (blue line) and b) (red line). The volume fraction of the crystal is plotted as a function of strain transmission, binned into 1% strain transmission intervals. The average strain delivered to the crystal is $\overline{\epsilon_{aa}} = 0.88\epsilon_{xx}^{disp}$ for model a), and $\overline{\epsilon_{aa}} = 0.67\epsilon_{xx}^{disp}$ for model b). Model b) has a wider volume fraction distribution of strain compared to direct gluing on the titanium.

[1]K. Behnia, *Fundamentals of thermoelectricity* (Oxford University Press, Oxford, 2015).

[2]C. Gayner and K. K. Kar Progress in Materials Science **83,** (2016).

[3]X. Shi, L. Chen and C. Uher International Materials Reviews **61,** (2016).

[4]G. Tan, L.-D. Zhao and M. G. Kanatzidis Chemical Reviews **116,** (2016).

[5]A. Stern, M. Dzero, V. M. Galitski, Z. Fisk and J. Xia Nature Materials **16,** (2017).

[6]J. Mutch, W.-C. Chen, P. Went, T. Qian, I. Zaky Wilson, A. Andreev, C.-C. Chen and J.-H. Chu eprint arXiv:1808.07898**,** (2018).

[7]J.-H. Chu, H.-H. Kuo, J. G. Analytis and I. R. Fisher Science **337,** (2012).

[8]C. W. Hicks, D. O. Brodsky, E. A. Yelland, A. S. Gibbs, J. A. N. Bruin, M. E. Barber, S. D. Edkins, K. Nishimura, S. Yonezawa, Y. Maeno and A. P. Mackenzie Science **344,** (2014).

[9]J. Park, H. Sakai, O. Erten, A. P. Mackenzie and C. W. Hicks Physical Review B **97,** (2018).

[10]T. Kissikov, R. Sarkar, M. Lawson, B. T. Bush, E. I. Timmons, M. A. Tanatar, R. Prozorov, S. L. Bud'ko, P. C. Canfield, R. M. Fernandes and N. J. Curro Nature Communications **9,** (2018).

[11]M. Matusiak, M. Babij and T. Wolf Physical Review B **97,** (2018).

[12]M. Matusiak, K. Rogacki and T. Wolf Physical Review B **97,** (2018).

[13]C. W. Hicks, M. E. Barber, S. D. Edkins, D. O. Brodsky and A. P. Mackenzie Rev Sci Instrum **85,** (2014).

[14]M. Eundeok, L. B. k. Sergey, S. T. Milton and C. C. Paul Measurement Science and Technology **21,** (2010).

[15]M. Meinero, F. Caglieris, G. Lamura, I. Pallecchi, A. Jost, U. Zeitler, S. Ishida, H. Eisaki and M. Putti Physical Review B **98,** (2018).

[16]E. D. Mun, S. L. Bud'ko, N. Ni, A. N. Thaler and P. C. Canfield Physical Review B **80,** (2009).

[17]S. Arsenijević, R. Gaál, A. S. Sefat, M. A. McGuire, B. C. Sales, D. Mandrus and L. Forró Physical Review B **84,** (2011).

[18]A. F. May, M. A. McGuire, J. E. Mitchell, A. S. Sefat and B. C. Sales Physical Review B **88,** (2013).

[19]Y. J. Yan, X. F. Wang, R. H. Liu, H. Chen, Y. L. Xie, J. J. Ying and X. H. Chen Physical Review B **81,** (2010).

[20]F. Caglieris, C. Wuttke, X. C. Hong, S. Sykora, R. Kappenberger, S. Aswartham, S. Wurmehl, B. Büchner, C. Hess arXiv:1905.11660**,** (2019).

[21]C. Fujii, S. Simayi, K. Sakano, C. Sasaki, M. Nakamura, Y. Nakanishi, K. Kihou, M. Nakajima, C.-H. Lee, A. Iyo, H. Eisaki, S.-i. Uchida and M. Yoshizawa Journal of the Physical Society of Japan **87,** (2018).

[22]A. Jain, *Materials Project (https://doi.org/10.1063/1.4812323, accessed 2019),* (2013).